\documentclass[amsmath,amssymb,twocolumn,amsfonts]{revtex4-2}
\usepackage{graphicx}
\usepackage{csquotes}
\def \beq {\begin{equation}}
\def \eeq {\end{equation}}
\def \ba {\begin{align}}
\def \ea {\end{align}}
\usepackage{bm}
\usepackage{bbm,mathbbol}
\usepackage{braket,bigints}
\usepackage{stmaryrd,mathtools}

\begin{document}
\title{Quantum thermodynamic derivation of the energy resolution limit in magnetometry}
\author{Iannis K. Kominis} 
\affiliation{Department of Physics, University of Crete, Heraklion 70013, Greece}
\begin{abstract}
It was recently demonstrated that a multitude of realizations of several magnetic sensing technologies satisfy the energy resolution limit, which connects a quantity composed by the variance of the magnetic field estimate, the sensor volume and the measurement time, and having units of action, with $\hbar$. A first-principles derivation of this limit is still elusive. We here present such a derivation based on quantum thermodynamic arguments. We show that the energy resolution limit is a result of quantum thermodynamic work necessarily associated with quantum measurement and Landauer erasure, the work being exchanged with the magnetic field. We apply these considerations to atomic magnetometers, diamond magnetometers, and SQUIDs, spanning an energy resolution limit from $10^0\hbar$ to $10^7\hbar$. This connection between quantum thermodynamics and magnetometry can help advance quantum sensing technologies towards even more sensitive devices.
\end{abstract}
\maketitle 
Magnetic fields convey useful information in diverse physical settings, therefore it is no surprise that quantum sensing \cite{Degen} of magnetic fields is one of the pillars of the second quantum revolution \cite{Deutsch}. Whether classical or quantum, magnetometers are characterized by several figures of merit like bandwidth \cite{band1,band2,band3,band4}, dynamic range \cite{dr1,dr2,dr3,dr4}, sensor size and scalability \cite{size1,size2,size3,size4}, accuracy \cite{acc1}, or the ability to operate in harsh environments \cite{harsh1,harsh2}. 

Magnetic sensitivity stands out as a prominent sensor characteristic, since the resolution of any measurement is limited by the intrinsic noise of the sensor. Advances in magnetic sensitivity brought about by superconducting sensors \cite{squid1,squid2} and optical pumping magnetometers \cite{op1,op2} spurred numerous applications, like sensing magnetic fields produced by the human brain \cite{meg1,meg2,meg3,meg4,meg5} or heart \cite{mcg1,mcg2}, materials characterization \cite{materials1,materials2}, even table-top probes of new physics \cite{fund1,fund2,fund3,fund4,fund5}. Therefore, understanding the fundamental limitations to magnetic sensitivity is crucial for pushing magnetic sensors towards optimal performance, and thus unraveling new applications.

By analyzing a large body of published work \cite{MitchellRMP}, it was demonstrated that tens of different realizations of several magnetic sensing technologies appear to have a unifying property, namely they all seem to satisfy the so-called energy resolution limit \cite{Robbes}. This limit states that $(\delta B)^2V\tau/2\mu_0\gtrapprox\hbar$, where $(\delta B)^2$ is the variance of the magnetic field estimate, $V$ the sensor volume, $\tau$ the measurement time, and $\mu_0$ the magnetic permeability of vacuum. The left-hand side of this bound has units of action. The fact that the right-hand side roughly equals $\hbar$ is aesthetically pleasing when discussing a fundamental limit. Surprisingly, as the authors in \cite{MitchellRMP} pointed out, a first-principles derivation of the energy resolution limit (ERL) is still elusive.

The ERL contains the expression $(\delta B)^2V/2\mu_0$, reminiscent of the magnetic energy within the sensor volume $V$. However, $(\delta B)^2/2\mu_0$ is not the actual magnetic energy density, since it contains $(\delta B)^2$ instead of $B^2$. Various attempts \cite{MitchellRMP} to derive the ERL based on quantum speed limits or energy-time uncertainty relations do not work, since instead of $(\delta B)^2$,  they involve the expression $\delta(B^2)=2B\delta B$, further leading to the counterintuitive result that $\delta B$ is suppressed when increasing $B$.

We will here derive the ERL based on quantum thermodynamic arguments. In particular, we will connect the ERL to the quantum work performed during quantum measurement and/or Landauer erasure of information. To this end, we treat the magnetic field as an integral part of the quantum thermodynamic environment of the sensor. When the quantum thermodynamic work accompanying the process of measurement is exchanged with the magnetic field energy, it leads to magnetic field fluctuations. 

The field of quantum thermodynamics \cite{Mahler, Kosloff, Parrondo, AndersCP, Goold,Deffner,Ozgur} has unified quantum information and quantum measurements with thermodynamic processes. The understanding of the physical nature of information \cite{Landauer, Bennett}, and the energy cost of information erasure \cite{Plenio} greatly inspired the development of quantum information science. Few works, however, have so far considered the connection of quantum thermodynamics with quantum metrology \cite{Anders2016,Lipka,Dutta,Cai}. The current work falls in this direction. 

The idea that the magnetic or electric field itself can act as a source/sink of quantum thermodynamic work is not new \cite{Anders2016,Koski}. We here push this idea further, towards understanding the ERL in magnetometry in a general way independent of the specific technology realization. The crux of the matter is the following. Let $u_B=B^2/2\mu_0$ be the magnetic energy density. If the field fluctuates by $\delta B$, where $\braket{\delta B}=0$, then the field energy within the volume $V$ of the sensor will change by $V\braket{u_{B+\delta B}-u_B}=(\delta B)^2V/2\mu_0$. If the cause of this fluctuation is the exchange of work $W$ between sensor and field, it will be $W=V\braket{u_{B+\delta B}-u_B}$, and the corresponding field fluctuation will be $\delta B=\sqrt{2\mu_0 W/V}$. Finally, using the relation $W=(\delta B)^2V/2\mu_0$ and quantum speed limits to connect the exchanged work $W$ with the time during which the exchange takes place, we will arrive at the ERL.

To proceed formally and find $W$ we use the approach of \cite{Ueda}, which establishes the minimum energy cost of measurement and Landauer erasure. The authors consider a system ${\cal S}$, a meter ${\cal M}$, and a thermal bath ${\cal B}$. A measurement is performed on ${\cal S}$, meaning that ${\cal S}$ and ${\cal M}$ become entangled. Then follows a projective measurement on ${\cal M}$. Finally, to make the process cyclic, the information in ${\cal S+M}$ is erased. The authors show that the combined work cost (done by ${\cal M}$ on ${\cal B}$) of measurement and erasure is bounded below: 
\beq
W_{\rm meas}+W_{\rm eras}\geq k_BT{\cal I},\label{SU}
\eeq
where $T$ is the bath temperature, with which the meter is in thermal contact, and ${\cal I}$ the mutual information between ${\cal S}$ and ${\cal M}$. This information satisfies $0\leq {\cal I}\leq H$, where $H$ is the Shannon entropy of the possible measurement results. If this work is exchanged between sensor and magnetic field, it will be $(\delta B)^2V/2\mu_0\geq k_BT{\cal I}$. If the exchange takes place during time $\tau$, we can use the Margolus-Levitin quantum speed limit \cite{ML,DC}, $k_BT{\cal I}\tau\geq {\pi\over 2}\hbar$, and finally we arrive at the ERL:
\beq
{{(\delta B)^2V\tau}\over {2\mu_0}}\geq{\pi\over 2}\hbar\label{erl}
\eeq
To apply the speed limit for the exchanged work the magnetic field is treated not as an external parameter, but as a physical degree of freedom. This can be done by e.g. quantizing the field (like in \cite{MitchellRMP}), and considering transitions moving energy $k_BT{\cal I}$ from states where this energy is localized in the sensor degrees of freedom, to orthogonal states where it is localized in the field. 

We will now specify the above for the case of optical pumping magnetometers, so the system ${\cal S}$ doing the sensing is an atomic vapor of spin-1/2 atoms (hyperfine structure is not relevant in this discussion). In the standard magnetometry framework the atomic spins are first spin-polarized by an optical pumping pulse, then precess due to the magnetic field, and finally are probed e.g. by a light beam. The magnetic field $B$ points along the $z$-axis, and the eigenstates of $\sigma_z$ are the computational basis states $\ket{0}$ and $\ket{1}$. The atoms are spin-polarized along the $x$-axis, their initial state being $\ket{\psi_0}=(\ket{0}+\ket{1})/\sqrt{2}$. The magnetic field evolves $\ket{\psi_0}$ into $\ket{\psi_\chi}=(\ket{0}+e^{i\chi}\ket{1})/\sqrt{2}$, where $\chi\propto B$. A measurement in the eigenbasis of $\sigma_x$ conveys information about $B$. 

However, in our analysis, as measurement in the quantum thermodynamic context of \cite{Ueda} we consider the dephasing produced by atomic collisions. In particular, in the SERF regime of interest here, spin destruction (SD) collisions \cite{Allred} are the fundamental mechanism for spin decoherence. In a binary SD collision one atom is the system ${\cal S}$ and another atom is the meter ${\cal M}$. During the collision, the two atomic spins become entangled, and further SD collisions act as a projective measurement on the meter atom, along the same lines described in \cite{MoulouPRA} for spin-exchange collisions. Thus, SD collisions can be seen as performing an unobserved measurement of $\ket{\psi_\chi}$ in the computational basis, pushing $\ket{\psi_\chi}$ towards $\rho={1\over 2}\ket{0}\bra{0}+{1\over 2}\ket{1}\bra{1}$. The measurement time $\tau$ is the duration of this process, i.e. the SD spin-relaxation time.

Information erasure (e.g. by an optical pumping pulse) renders the whole process cyclic. Nevertheless, information erasure does not incur an energy cost, since the optical pumping photons are scattered into the light field having practically zero temperature \cite{Rochester}. Hence in our case, the thermodynamic work entering \eqref{SU} is solely due to the aforementioned measurement. 

The Shannon entropy of the decohered state $\rho$ is $H=\ln 2$. We still need to determine the mutual information ${\cal I}$. When a system atom collides with a meter atom, their spins become entangled. The entanglement is not maximal, because the atom's spin phase change in SD collisions is small \cite{Walker}, unlike the case of spin-exchange collisions \cite{primer}. Hence it might appear that ${\cal I}\ll H$. However, after many such binary collisions, the system state is fully decohered. We consider all those collisions as one process mapping $\ket{\psi_\chi}$ into $\rho$ along the measurement time $\tau$, during which the total mutual information ${\cal I}=H$. It is interesting to note that this information of 1 bit pertains to the whole vapor, as individual atoms cannot be addressed. Experiments with hot vapors measure the ensemble spin, and (apart from relaxation effects attributed to binary collisions) the measurements reflect single-atom information, the signals being amplified by $N$, the number of atoms.

To apply the bound $k_BT{\cal I}\tau\geq {\pi\over 2}\hbar$, we crucially note that the temperature $T$ should be the spin temperature $T_s$. While the translational degrees of freedom are indeed governed by the thermodynamic temperature, which is the room temperature or higher, the spin temperature can be widely different. It is known that spin-exchange collisions lead to a spin-temperature distribution \cite{primer,Appelt}. Moreover, at low spin polarization pertinent to understanding sensor noise, spin-destruction collisions obey the same spin dynamics as spin-exchange collisions, and thus also lead to a spin-temperature distribution \cite{primer,Appelt}. Without optical pumping, the spin-temperature at equilibrium is infinite (diagonal density matrix). However, it is known from spin-noise spectroscopy \cite{SN1,PRA2007,SN2,SN3,SN4} that there are spontaneous spin-polarization fluctuations of order $1/\sqrt{N}$ around this equilibrium, hence $T_s$ is finite.

To find $T_s$ when the magnetic field is $B$ we relate $\mu B/2k_BT_s$, where $\mu$ the atom's magnetic moment, with the spin-noise polarization of the vapor along the $z$ axis, which is on the order of $1/\sqrt{N}$. Consider the spin-temperature density matrix $\rho=e^{-H/k_BT_s}/{\rm Tr}\{e^{-H/k_BT_s}\}$, where the magnetic Hamiltonian is $H=-{{\mu B}\over 2}\sigma_z$. Setting $2k_BT_s=\mu\sqrt{N}B$, and expanding $\rho\approx{1\over 2}(1+\sigma_z/\sqrt{N})$, it follows that indeed $\braket{\sigma_z}={\rm Tr}\{\rho\sigma_z\}\approx 1/\sqrt{N}$.

This point is further substantiated: an atom in the state $\ket{\psi_{\chi}}$ has zero energy, given that $H\propto\sigma_z$. When the atom's state is finally projected to either $\ket{0}$ or $\ket{1}$ by SD collisions, the field provides the atom with energy $\pm \mu B/2$. Projection to either $\ket{0}$ or $\ket{1}$ has probability 1/2, thus the total energy exchanged with the field is on average zero, apart from an uncertainty $\mu B\sqrt{N}/2$ (binomial distribution with equal probabilities). This is the same quantity derived previously as $k_BT_s$. 

Finally, this result can be generalized by considering the time-dependent relaxation dynamics of the vapor, starting from the fully spin-polarized state $\ket{\psi_\chi}=(\ket{0}+e^{i\chi}\ket{1})/\sqrt{2}$. For this state, the spin polarization along $\mathbf{\hat z}$, given by $\braket{\sigma_z}$, is exactly zero. During the relaxation transient, we can write $d\braket{\sigma_z}_t=(1-|\braket{\sigma_+}_t|)d\xi$, where $d\xi$ is a Gaussian noise process of variance $1/N\tau$, and $|\sigma_+|=e^{-t/\tau}$ is the transverse spin decaying with time constant $\tau$. That is, the spin polarization along $\mathbf{\hat z}$ is described by a nonlinear noise term, reflecting its dependence on the spin state along the transient, during which atoms are {\it gradually} projected from the sate $\ket{\psi_\chi}$ to $\ket{0}$ or $\ket{1}$. After $|\sigma_+|$ has decayed to zero, SD collisions still drive spin fluctuations along $\mathbf{\hat z}$, thus $d\braket{\sigma_z}_{t\gg\tau}=d\xi$. The average variance of $\braket{\sigma_z}_t$ during the relaxation time $\tau$ is given by ${1\over {N\tau}}\int_0^\tau(1-e^{-t/\tau})^2dt$, which translates to an average uncertainty $\delta\braket{\sigma_z}\approx 0.5/\sqrt{N}$. In other words, it is the late time noise that dominates the result, which apart from the factor of 0.5, was found in the main text by considering the equilibrium spin-temperature state instead of the relaxation transient. It is noted that fluctuations in $\braket{\sigma_z}$ due to external probing of the spins are in addition to those considered here, which are due to the \enquote{internal} measurements due to SD collisions. An optimized probing, however, would be designed so as to not significantly decrease the relaxation time $\tau$, so our result is approximately valid even in this more general case. Overall, 
\beq
k_BT_s=\mu\sqrt{N}B\label{Ts}
\eeq
To proceed, we set $B\approx\delta B$, i.e. work at small magnetic fields close to the sought after intrinsic sensor noise. Parenthetically, it would be interesting to investigate the ERL at higher magnetic fields. There we anticipate rich phenomenology, since higher magnetic fields produced by degrees of freedom external to the sensor (e.g. a current-carrying coil) might open new channels for the exchange of thermodynamic work between the sensor and its environment. Thus the relation $(\delta B)^2V/2\mu_0\geq k_BT{\cal I}$, which formally leads to $\delta B\propto\sqrt{B}$ in this discussion of atomic sensors, cannot be simply extrapolated to arbitrarily high fields. 

Using the bound $k_BT_s{\cal I}\tau\geq \pi\hbar/2$, we find $\delta B\geq (\pi/2\ln(2))(\hbar/\mu\sqrt{N}\tau)$. The relaxation time due to SD collisions is given by $1/\tau=n\sigma_{\rm sd}\overline{v}$, where is $n=N/V$ the atom number density, $\sigma_{\rm sd}$ the SD cross section, and $\overline{v}$ the relative velocity of the colliding partners. Hence
\beq
\delta B\geq{\pi\over {2\ln2}}{{\hbar\sigma_{\rm sd}\overline{v}\sqrt{N}}\over {\mu V}}\label{dBN}
\eeq
Using \eqref{dBN} we can finally obtain the ERL, which reads
\beq
{{(\delta B)^2V\tau}\over {2\mu_0}}\geq\hbar\Big({{\pi^2}\over {8(\ln 2)^2}}{{\hbar\sigma_{\rm sd}\overline{v}}\over {\mu_0\mu^2}}\Big)\label{erl2}
\eeq
The atom's magnetic moment is $\mu=\mu_B/q$, where $\mu_B$ is the Bohr magneton and $q=[(S(S+1)+I(I+1)]/S(S+1)$ is the nuclear slowing down factor with $S=1/2$ and $I$ the nuclear spin \cite{Allred}. Using the SD cross sections \cite{Allred} and the respective relative velocities $\overline{v}$, a number density $n=10^{14}~{\rm cm^{-3}}$ and volume $V=10~{\rm cm^{3}}$, we list in Table I the values of $\delta B$ of \eqref{dBN} and the respective ERL of \eqref{erl2} for $^{41}{\rm K}$, $^{87}{\rm Rb}$ and $^{133}{\rm Cs}$.
\begin{table}[t!]
\caption{Atomic Magnetometer ERL. The numbers reflect the magnetic field noise and corresponding ERL for an atomic vapor of atom number density $10^{14}~{\rm cm^{-3}}$ and volume $10~{\rm cm^{3}}$.}
\begin{ruledtabular}
\begin{tabular}{|c|c|c|c|}
Atom                         & $^{41}{\rm K}$   &  $^{87}{\rm Rb}$         & $^{133}{\rm Cs}$ \\
\hline
$\delta B$  ($10^{-17}~{\rm T}$) &      2                  &  10      &     1000 \\
Predicted ERL ($\hbar$)   &         4           &    25        &     6054 \\
\end{tabular}
\end{ruledtabular}
\end{table}

For a comparison, the authors in \cite{RomalisPRA} used a cesium vapor of volume $V=1~{\rm cm^{3}}$ and number density $2\times 10^{13}~{\rm cm^{-3}}$, which by \eqref{dBN} lead to $\delta B\approx 10^{-14}~{\rm T}$. Since this noise is distributed within the bandwidth $1/\tau$, the corresponding spectral density is $10~{\rm pG/\sqrt{Hz}}$, whereas the authors measured a noise level of $400~{\rm pG/\sqrt{Hz}}$ and, under the premise of a theoretical optimization, project noise $2~{\rm pG/\sqrt{Hz}}$. 

We will now elaborate further on the bound \eqref{dBN}, rewriting it as 
\beq
\delta B\geq{2\over 3}{\kappa\mu_0\mu\sqrt{N}\over V},\label{dBNk}
\eeq
in order to exhibit the magnetic field produced within a spherical magnetized volume of a randomly spin-polarized vapor. Only now, this dipolar field is seen to be amplified by the factor $\kappa=({{3\pi}/{4\ln 2}}){{\hbar\sigma_{\rm sd}\overline{v}}/{\mu_0\mu^2}}$. 
On the one hand we can argue, like in \cite{MitchellRMP}, that this amplification is forced by the uncertainty relation. Given that the uncertainty $\Delta\sigma_x$ is negligible for a vapor optically pumped along the $x$-axis, the uncertainty $\Delta\sigma_y$ will be determined by the field $\delta B$ driving spin precession from the $x$ axis to the $y$-axis, i.e. $\Delta\sigma_y=\delta\phi\braket{\sigma_x}$ where $\delta\phi=\mu \delta B\tau/\hbar$. Since $\Delta\sigma_y$ and $\Delta\sigma_z$ have to satisfy the uncertainty relation $\Delta\sigma_y\Delta\sigma_z\geq|\braket{\sigma_x}|$, it follows by setting $\Delta\sigma_z=1/\sqrt{N}$ that $\delta B$ is given by \eqref{dBNk}. 

There is yet another interpretation for the factor $\kappa$, interesting in its own right. We can consider the vacuum permeability $\mu_0$ replaced by $\kappa\mu_0$, as if the spins have a tendency to align, and $\kappa$ represents the relative permeability constant. But indeed, binary collisions do correlate atoms, as was shown in \cite{MoulouPRA} for spin-exchange collisions. It was shown that the negativity of the two-atom spin state scales as $\sin\phi_{\rm se}^2$, where $\phi_{\rm se}\approx 1$ is the spin-exchange phase change. Spin-destruction collisions incur a spin phase change $\phi\ll 1$. Since at low spin polarization pertinent to the study of sensitivity limits (see also discussion in End Matter) the dynamics of spin-exchange and spin-destruction are the same \cite{Appelt}, we can write that the entanglement between two atoms after an SD collision will scale like $\sin\phi^2\approx\phi^2$, since now $\phi\ll 1$. Hence any two colliding partners will share a small correlation. The upside is that this correlation can be shared by many more atoms. Imagine a "bath" atom experiencing consecutive collisions with many "system" atoms. These will all be slightly correlated, and will reside in a "correlation volume" $V_c$. Consider two spins inside this volume, which are about to collide. Their interaction energy is $\epsilon=\kappa\mu_0\mu^2/\tilde{V}$, where $\tilde{V}$ is the volume defined by the two spins. This interaction reorients the spins in a time scale $\hbar/\epsilon$. For the aforementioned correlation to be maintained, this time should be equal or larger than $\tilde{V}/\sigma_{\rm sd}\overline{v}$, which is the time required for their collision to correlate them. From this requirement we obtain again $\kappa=\hbar\sigma_{\rm sd}\overline{v}/\mu_0\mu^2$. We can also estimate the correlation volume $V_c$, which will contain $N_c$ atoms. We can write $\kappa=\sqrt{N_c}\phi$, since the spin variance scales as $\phi^2$ and thus the uncertainty as $\phi$, while for $\phi\approx 1$ we would get an enhancement $\kappa=\sqrt{N_c}$, retrieving the case of fully polarized spins. The phase $\phi$ is related to the SD relaxation time and the collision rate $1/T$, i.e. it is $1/\tau=\phi^2/T$ as discussed in \cite{primer}. The collision rate is $1/T=\overline{v}/n^{-1/3}$. Putting everything together we find $N_c=(\hbar\overline{v}/\mu_0\mu^2)^2\sigma_{\rm sd}n^{-2/3}$. For a $^{41}{\rm K}$ number density of $n=10^{14}~{\rm cm}^{-3}$ it is $N_c\approx 10^9$, and $V_c=N_c/n\approx 0.01~{\rm mm}^3$. For $^{133}{\rm Cs}$ these numbers would be $N_c\approx 10^{13}$, and $V_c\approx 0.1~{\rm cm}^3$. 

We finally note that the correlation described by $\kappa$ is not inconsistent with a spin-temperature distribution. We can rescale $T_s$ as given by Eq. \eqref{Ts} by $\kappa$, i.e. $T_s\rightarrow T_s/\kappa$, and immediately arrive at expression \eqref{dBNk} by considering the magnetic field produced by the noise polarization of this vapor, which will now read $\braket{\sigma_z}\approx\kappa/\sqrt{N}$. Stated differently, this correlation embodied in $\kappa$ reflects a slightly cooler spin temperature, or equivalently, a higher noise polarization. 

The physical picture that emerges is that the atomic vapor will behave like a \enquote{squashy} spin medium exhibiting spin fluctuations, split in a number of \enquote{domains} not unlike ferromagnetic systems. Such domains exhibit intra-domain correlations, but not inter-domain correlation. Their contributions to the noise field add in quadrature, hence the total noise field is still given by \eqref{dBNk}. The vapor as a whole continuously exchanges energy with its self-field, setting an unavoidable noise level for measuring externally applied magnetic fields. Towards an experimental verification of the noise $\delta B$, we propose the use of a miniaturized cesium cell of volume e.g. $1~{\rm mm^{3}}$, in order to boost the previous estimates of $\delta B$ by a factor of $10^4$. To avoid wall relaxation the cell should have buffer gas in order to avoid wall relaxation \cite{Kitching,Hasegawa}, or  anti-relaxation coating \cite{Rooij}. Also, the cell should be surrounded by a superconducting flux transformer in order to alleviate Johnson noise that would dominate the signal induced by the changing flux produced by $\delta B$ in an ordinary coil.
 
The possibility that spin uncertainty in an atomic vapor creates a fluctuating field consistent with the ERL was discussed in \cite{MitchellRMP} and references therein, but largely dismissed, mainly on grounds of angular momentum and energy conservation. The physical picture painted here differs in several subtle ways from \cite{MitchellRMP}: (i) We claim that (a) when starting from a spin-polarized state, which during the time $\tau$ relaxes to the equilibrium state exhibiting spin-noise, energy will be exchanged between atomic spins and magnetic field, and (b) this process will continue even if the atoms are left alone in their equilibrium state. While the authors in \cite{MitchellRMP} refer to quantum-measurement induced spin uncertainty, we refer to spontaneous spin noise that exists in such vapors whether there is or isn't an external measurement limited by quantum uncertainty. (ii)  There is no issue with angular momentum conservation, since in SD collisions angular momentum is taken up by translational angular momentum of the colliding atoms. Similarly, there is no issue with energy conservation, since spin dynamics are driven by collisions, the translational energy of which is supplied by an external energy source. A tiny fraction of this energy is transformed into magnetic interactions through SD and other spin-dependent collisions. (iii) The authors in \cite{MitchellRMP} postulated such magnetic field fluctuations emanating from spin uncertainty, and showed they are consistent with the ERL. We started from the opposite direction, postulating the thermodynamic work exchanged between sensor and field, and arriving at these spin fluctuations and their self-interaction, connecting the thermodynamic work to the specific measurement process going on during SD spin relaxation. (iv) The authors in \cite{MitchellRMP} considered two different noise sources, one stemming from the uncertainty relation discussed above, and another from the field produced by randomly polarized spins. Then they elaborated on the difficulties of adding those two terms in quadrature. We unify these into a single noise term interpreted with the novel type of correlations manifested as an enhanced permeability of the vapor. (v) The quantum speed limits did not work in the considerations of \cite{MitchellRMP}, because they concerned the total energy $B^2V/2\mu_0$. Here they concern the energy {\it exchanged} between atoms and field, which is $(\delta B)^2V/2\mu_0$.  

A pending question is whether the ERL is a hard fundamental limit. In fact, there are measurements claiming an ERL below $\hbar$ \cite{belowERL1,belowERL2,belowERL3}. We note that the magnetic field fluctuations in \eqref{dBN} were derived assuming a spin-temperature equilibrium state of the atomic vapor.  Such thermal equilibrium state might not be the case, thus a basic assumption of the derivation of the quantum thermodynamic work in \cite{Ueda} would break down in the first place. For example, for a spin-squeezed equilibrium state, the spin fluctuations would be suppressed by some squeezing factor $\xi\leq 1$ \cite{squeezing}. Correspondingly, the ERL in \eqref{erl2} would be violated by the factor $\xi^2$. In this respect, the claims  in \cite{belowERL1,belowERL2,belowERL3} seem plausible. Given that the Heisenberg limit allows $\xi^2=1/N$, where $N$ is the atom number, one could imagine an ultra-low ERL of $\hbar/N$. This conclusion tacitly assumes that the relaxation time $\tau$ is not $\xi$-dependent. However, the physics of spin relaxation in correlated vapors is far from being understood \cite{BudkerPRL,KominisPRL,MitchellNC}. Thus, notwithstanding the claims \cite{belowERL1,belowERL2,belowERL3}, we cannot make strong statements about the general validity of the ERL.

We will, however, provide two more demonstrations of the quantum thermodynamic ERL, for the SQUID and the diamond sensor \cite{D1,D2}. Going back to the relation $(\delta B)^2V/2\mu_0=k_BT{\cal I}$, these sensors should satisfy $k_BT{\cal I}\tau\geq{\pi\over 2}\hbar$. Regarding diamond sensors, the electron spin relaxes by the end of the measurement, hence as in the case of atomic vapors, the extracted information is ${\cal I}=\ln 2$. We use the numbers of \cite{Patel}, where a room-temperature diamond sensor has spin relaxation time $\tau\approx 1~{\rm \mu s}$. We find an optimal ERL of $k_BT{\cal I}\tau\approx 3\times 10^7\hbar$. The magnetic sensitivity reported in \cite{Patel} is $\delta B=300~{\rm pT}/\sqrt{\rm Hz}$, and the measurement volume estimated as $V=600~{\rm \mu m\times 4650~\mu m^2}$. Thus, the measured ERL reads $10^9\hbar$, so a 30-fold improvement is possible. The authors in \cite{Patel} state $\delta B$ is a factor of 6 away from the photon shot-noise limit. If this saturates the ERL, the sensing volume would need to be revised by 20\%. In any case, such considerations illuminate the constructive role the quantum-thermodynamic ERL can play in understanding and optimizing magnetic sensors.

For SQUIDs the temperature $T$ is on the order of $1~^\circ~{\rm K}$. What is the information ${\cal I}$ in this case? The SQUID could be seen as distinguishing between a flux $n\Phi_0$ and a flux $(n+1)\Phi_0$, where $\Phi_0$ is the flux quantum. 
\begin{table}[t!]
\caption{SQUID ERL. Comparison between measurements and quantum thermodynamic prediction}
\begin{ruledtabular}
\begin{tabular}{|c|c|c|c|c|c|}
Publication                         &  \cite{Schmelz2017}    &  \cite{Wakai}          &  \cite{Awschaiom} & \cite{Schmelz2016} & \cite{Schmelz2011}\\
\hline
$p$  ($10^{-6}$)                         &  0.045         &      5.5   & 0.084  &  0.6   & 1.3\\
$T ({\rm ^\circ K}$)             &        4.2                &     1.5                 &  0.29                       &        4.2                         & 4.2\\
$\tau$ ($10^{-5}~{\rm s}$) &             0.5             &     0.013                                     &       5.0              &        0.5                       &        0.5  \\
Predicted ERL ($\hbar$)   &         2.1              &    1.6                 &     2.6             &      24                   & 50\\\hline
Measured ERL ($\hbar$)   &        6.3              &    1.6                 &     1.7            &       35                   & 121
\end{tabular}
\end{ruledtabular}
\end{table}
These two altrenatives would indeed correspond to 1 bit of information. However, in the flux-locked-loop operating mode \cite{squid1,squid2}, the actual flux noise (the excursions of the flux around the operating point $(n+1/2)\Phi_0$) is on the order of $p\Phi_0$, hence the gained information is ${\cal I}\approx-p\ln p$, where $p\ll 1$. We thus expect $(-p\ln p)k_B T\tau\geq{\pi\over 2}\hbar$. To demonstrate this, we use the results of five SQUID papers \cite{Schmelz2017,Wakai,Awschaiom,Schmelz2016,Schmelz2011} reporting (i) the flux noise as a fraction of $\Phi_0$ per ${\rm Hz}^{1/2}$, from which we extract the parameter $p$ given a 1 Hz resolution bandwidth, and (ii) the measurement bandwidth, from which we extract $\tau$. The authors in \cite{Schmelz2017,Wakai,Awschaiom,Schmelz2016,Schmelz2011} explicitly report the measured ERL, with which we compare the predicted ERL, i.e. the expression $(-p\ln p)k_B T\tau$. It is seen at Table II that the measured ERL is higher and of the same order compared to the quantum thermodynamic ERL (except the case of \cite{Awschaiom}, where they coincide within the experimental errors of \cite{Awschaiom}). Thus regarding SQUIDs, the quantum thermodynamic ERL appears to work for a range from $1\hbar-100\hbar$. 

Concluding, we have presented a first-principles derivation for the energy resolution limit in magnetic sensing, specified for atomic spin magnetometers, diamond sensors and SQUIDs. We demonstrate the applicability of the ERL for starkly different physical parameter values relevant to three starkly different sensor technologies, over a range of seven orders of magnitude in $\hbar$. The understanding of the physical principles behind the ERL could have multifaceted consequences for the effort to advance quantum sensing technologies.

\end{document}